\documentstyle[aps,twocolumn,prl,epsf]{revtex}

\begin{document}

\draft

\title{Size-Dependent Transition to High-Dimensional Chaotic Dynamics in a Two-Dimensional Excitable Medium}

\author{
Matthew~C.\ Strain\cite{CNCS-address}\cite{strain-email} and
Henry~S.\ Greenside\cite{CNCS-address}
}

\address{
Department of Physics\\
Duke University, Durham, NC 27708-0305
}

\date{January 6, 1997}

\maketitle


\begin{abstract}
The spatiotemporal dynamics of an excitable medium with
multiple spiral defects is shown to vary smoothly with
system size from short-lived
transients for small systems to
extensive chaos for large systems.  
A comparison of the Lyapunov dimension 
density with the average spiral defect
density suggests an average dimension per spiral defect 
varying between three and seven.  
We discuss some implications of these results
for experimental studies of excitable media.
\end{abstract}


\narrowtext
Much research in the nonequilibrium physics of excitable
media has been motivated by the observation of dynamical
states containing defects, {\it i.~e.}
spiral waves in two space dimensions
or spiral filaments in three dimensions~\cite{spiral-refs}.
Experimental studies 
in surface oxidation experiments~\cite{co-exper-refs} 
and 
in fibrillating hearts~\cite{heart-refs}
suggest that many such defects may coexist in dynamically complex states.
Although similar states have been reproduced
in computer simulations~\cite{baer-refs,Karma,Winfree91}, 
there has not yet been careful
quantitative analysis of whether the long-time dynamics
of such media can be chaotic and, if so, how the properties
of this chaos may be related to the statistics of defects,
to the size of the medium, and to intrinsic medium parameters.
Detailed analysis of mathematical models of 
excitable media may thus provide new insights in how to
analyze spatially-extended excitable media, 
possibly including fibrillating cardiac tissue.

In this Letter, we numerically study a
two-dimensional model of a homogeneous excitable
medium with an 
emphasis on determining
when spatiotemporal chaos occurs, and on
quantitatively analyzing basic time and length scales of
observed chaotic states.  
We use a model introduced by B\"ar {\it et~al.}~\cite{baer-refs}, 
as a reduced description of carbon monoxide oxidation on a
surface~\cite{Baer94jcp}, because of its numerical simplicity
and because of prior work suggesting the existence of chaos
\cite{Hildebrand95}.
Extending some recent work by other researchers
\cite{baer-refs,Karma,Panfilov95}, we find that the dynamics
is strongly dependent on the system size $L$.  
For small~$L$, all initial conditions studied rapidly
decay to an asymptotic constant or periodic state.
As the system size increases, however,
the volume of the set of initial conditions leading 
to sustained, non-periodic dynamics 
increases smoothly, and we discuss the transition
from periodic to non-periodic dynamics with increasing
system size.
The non-periodic dynamics sustained in sufficiently
large systems are statistically stationary, and	 
we compute Lyapunov exponents and dimensions, 
defect statistics,
and two-point correlation lengths to characterize these states.
These different statistics are compared,
both to test previous conjectures about the relationship
of these correlation lengths~\cite{Bayly93}
and to evaluate the complexity of the defects.
Our results indicate that a two-dimensional excitable medium of
moderate size with few defects on average 
can already sustain extensive, high-dimensional chaotic dynamics,
a fact with important implications for control of excitable
media by small parameter perturbations~\cite{control-refs}.
In the following, we explain the model, 
summarize our calculations, and discuss the implications 
of our results.

The B\"ar model describes the 
interaction of an activator field $u(t,x,y)$
with an inhibitor field $v(t,x,y)$ via the partial
differential equations
\begin{mathletters}
\label{eq:bar-model}
\begin{eqnarray}
  { \partial{u} \over \partial{t} }
    &=&  \nabla^2{u}
       +  \frac{1}{\epsilon} u (1 - u) \left(u -
       \frac{v+b}{a} \right) , \label{eq:u-eq} \\ 
  { \partial{v} \over \partial{t} }
     &=& f(u) - v ,\label{eq:v-eq} \\
\end{eqnarray}
\end{mathletters}
which we solve numerically in a square domain of side~$L$
with either biperiodic (BP) or no-flux (NF) 
boundary conditions on the field~$u(t,x,y)$.  The
function~$f(u)$ has the form
\begin{equation}
 f(u)=
    \left\{ \begin{array}{ll}
     0 ,              & \mbox{if $u \le 1/3$} ,\\
     1-6.75u(u-1)^2 , & \mbox{if $1/3 \le u \le 1$} ,\\
     1 ,              & \mbox{if $u > 1$} ,
            \end{array}
     \right. 
\label{eq:f-defn} 
\end{equation} 
so that the production of the inhibitor $v$
is ``delayed'' until $u$ exceeds $1/3$.
The nonlinear form Eq.~\ref{eq:f-defn}~leads to three
fixed points, one stable and two unstable; the larger
unstable fixed point $(u^*,v^*)$
(which does not appear in the 
widely-used Fitzhugh-Nagumo model)
seems necessary for the occurrence of spatiotemporal
chaos. 
The parameter~$\epsilon$ in Eq.~\ref{eq:u-eq}
determines the ratio of time scales of the fast field~$u$
and slow field~$v$ and is the key bifurcation parameter in
this paper. The positive parameters~$a$ and~$b$ were fixed
at the values $a=0.84$ and $b=0.07$ to take advantage of
substantial earlier work using these
values~\cite{baer-refs,Hildebrand95}.  
Spiral solutions are then known
empirically to be unstable when~$\epsilon$ exceeds a
critical value $\epsilon_c \approx 0.069$~\cite{baer-refs}.
The mechanism of this instability, meander of the spiral
core into a branch of the spiral, is apparently unique
to models with ``delayed'' inhibitor production 
like that given by Eq. (\ref{eq:f-defn}).
In particular, this is not the mechanism of 
breakup observed in models of cardiac 
tissue~\cite{Karma,Winfree91}.
Breakup leads to long-lived, complicated
dynamics for certain initial conditions when~$\epsilon >
\epsilon_c$; a snapshot of such a disordered nonperiodic
state with~$31$ spiral defects is shown
in Figure 1.

Our calculations involved integrating Eq. (\ref{eq:bar-model}),
calculating the Lyapunov spectrum of the numerical
trajectory, and counting the number of spiral defects at
successive times. For both kinds of boundary conditions,
Eq. (\ref{eq:bar-model}) was solved numerically by first
introducing second-order accurate finite difference
approximations for the spatial derivatives on a uniform
square mesh of spacing~$\triangle{x}$ and then using an
algorithm proposed by
Barkley~\cite{Barkley91}.  For the calculations reported
below, we used a spatial grid size $\triangle x=0.50$ and
time step $\triangle{t} = 0.05
\epsilon$. The spectrum of Lyapunov
exponents~$\lambda_i$ and the Lyapunov fractal dimension~$D$
were calculated by well-known algorithms based on linear
variational equations~\cite{Parker89} that were integrated
by a forward-Euler algorithm with the same grid and time
step.  The time step chosen, $\triangle{t}= 0.05\epsilon$,
was much smaller than that required by integration of only
Eq. (\ref{eq:bar-model}), but was
found to be necessary to compute Lyapunov 
exponents accurately to within a few percent~\cite{strain-ms-thesis97}.
For given boundary conditions and initial data,
Eq. (\ref{eq:bar-model}) was integrated for~$2000$ time units to
allow a statistically stationary state to be obtained,
and then the full system with variational equations was
integrated for an additional $1000$~time units
($\approx 200$ spiral periods), during which statistics
were calculated.

To study the dependence of the dynamics on initial conditions,
we integrated Eq. (\ref{eq:bar-model}) from each of 
$100$ initial conditions 
generated by distributing the field values uniformly 
(at each grid point) in the ranges
$u \in [0.8 u^*, 1.2 u^*], v \in [0.8 v^*, 1.2
v^*]$.  This procedure was repeated for both boundary conditions
and for square systems of side length $L$ varying from 5 to 40.
For all initial conditions,
the dynamics was short-lived in small systems
($L < 15$ (for NF boundary condition)  or $L<8$ (for BP) ), 
decaying in less than~$100$ time units to either 
the stable uniform state or to a plane-wave
state (only in the case of biperiodic boundary conditions).
Sufficiently large systems ($L > 35$)
sustained dynamics for at least $3000$ time units. 
The fraction $f$ of initial conditions which led to
non-periodic dynamics sustained for a time $T$ 
(either 100 or 1000 time units)
is shown in Figure~\ref{fig:transients}
as a function of system size for both boundary conditions.
For biperiodic boundary conditions (Figure 2(a)), the curve is
independent of the cutoff time $T$, indicating that
transients either die quickly or are sustained indefinitely 
(more than $50,000$ time units).  
For no-flux boundary conditions, all initial
conditions studied eventually decayed to a stationary state;
the mean and median transient times both scale 
exponentially with system size~\cite{strain-ms-thesis97},
much like the supertransient behavior observed previously
in one-dimensional systems~\cite{transient-refs}.
These results suggest that excitable media of intermediate
size may have an appreciable basin of attraction 
both for non-periodic dynamics and for periodic or constant
dynamics.
Whether this accounts for the observation
that fibrillation {\em sometimes} 
occurs in hearts of intermediate size remains unclear,
both because of the differing breakup mechanisms and
because of the effect of the third dimension in heart
tissue~\cite{Panfilov95,Winfree95}.

The fact that a given state was transient 
was revealed only by an eventual abrupt change
to the uniform state; the dynamics of the transient
itself was found to be statistically stationary.
For parameter values~$\epsilon > \epsilon_c$ and for system
sizes~$L > 25$, these statistically stationary states were
found to be high-dimensional ($D \ge 20$) and extensively
chaotic~\cite{Cross93} as shown in
Figure 3 by a linear
dependence of~$D$ on~$L^2$.  From the asymptotic slopes of
the curve in Figure 3, an intensive
dimension density~$\delta = \lim_{L^2 \to \infty}
\partial{D} / \partial{(L^2)}$ was obtained and then
re-expressed as a dimension correlation length $\xi_\delta =
\delta^{-1/d}$ for a ${d=2}$~dimensional
domain~\cite{Cross93}.  To test a speculation of Bayly {\em et. al.}
\cite{Bayly93} that knowledge of the
experimentally-accessible two-point correlation
length~$\xi_2$ might provide knowledge of the dynamical
length~$\xi_\delta$ for a chaotic state of spiral defects, 
we computed $\xi_2$
and~$\xi_\delta$ for several values of the
parameter~$\epsilon$. For each $\epsilon$~value studied, the
two-point correlation function~$C(r)$ had a similar
monotonically-decreasing but non-exponential form so we
estimated~$\xi_2$ by the position of the first zero crossing
of $C(r)$~\cite{strain-ms-thesis97}.  As shown in
Figure 4(a), the two lengths agree within a
factor of 1.5 or better but have opposing trends
as~$\epsilon$ increases (from $0.07$ to $0.12$), 
with $\xi_\delta$ decreasing 
and~$\xi_2$ increasing slightly.
It is unclear from this data whether an estimate
of $\xi_\delta$ can, in general, be obtained by
measuring $\xi_2$.
In any case, further analysis
of more physiologically accurate models will be needed to
relate~$\xi_\delta$ and~$\xi_2$ for the heart data of Bayly
{\em et. al.} \cite{Bayly93}.

We also explored whether the fractal dimension~$D$ of the
chaotic states was related to the statistics of the
number~$N(t)$ of spiral defects, {\em e.g.}, to its time
average~$\langle N \rangle$. The spirals were counted at
successive times by locating their cores, which occur at
those points~$(x,y)$ in the medium where the fields~$(u,v)$
take on the values $(u^*,v^*)$~\cite{baer-refs}.  
It has been shown previously that $N(t)$
is constant for $\epsilon < \epsilon_c$, in which case the
time average $\langle N\rangle$ is fixed by the choice of
initial condition~\cite{baer-refs}.  
We found that the mean~$\langle
N \rangle$ scaled extensively with system area~$L^2$ for
both boundary conditions considered, so that the average
defect density $n = \langle N \rangle/L^2$ was independent
of system size~\cite{Hildebrand95}.
The ratio $d = D/{\langle N\rangle}$ of the Lyapunov
dimension to the mean number of defects therefore defines an
intensive quantity that measures the number of dynamical
degrees of freedom associated with each defect on average.
Because the extensive quantities $D$ and $\langle N
\rangle$ are of the form $\alpha L^2 + \beta$ 
rather than simply $\alpha L^2$ (where $\alpha$ and $\beta$
are constants), the ratio~$d$ asymptotes slowly to a constant
value close to $\delta/n$.  We studied the dependence of
$D/\langle N \rangle$ on area~$L^2$ for two
different values of~$\epsilon$, and found that we could
estimate the ratio~$d$ accurately using a single system of side
length~$L=40$ with periodic boundary conditions.  For this
system size, $d$~increases smoothly with~$\epsilon$ from
less than~$4$ for $\epsilon \approx \epsilon_c$ to nearly
$7$ for $\epsilon \geq 0.095$ at which point it becomes
approximately constant (see Figure 4(b)).  Thus a
fixed number of degrees of freedom can not, 
in this manner, be associated
with each spiral defect in an excitable medium. 

In summary, for a particular model of a two-dimensional
excitable medium~\cite{baer-refs,Hildebrand95}, we have 
demonstrated by
numerical calculations a smooth
transition from short-lived
transient dynamics to extensive,
high-dimensional chaotic dynamics 
with increasing system size~$L$ for both no-flux
and biperiodic boundary conditions.
Small systems never exhibited sustained chaotic dynamics,
but non-periodic dynamics in sufficiently large systems 
were found to be statistically stationary on such long time scales 
and for such a large fraction
of random initial conditions that Lyapunov 
spectra of the dynamics converged well to a value 
independent of the initial condition.  
Although previous results based on time series allowed
computation of one Lyapunov exponent 
\cite{ZhangHolden95,Garfinkel97},
we computed enough Lyapunov exponents 
to determine the Lyapunov dimension~$D$ in an excitable 
medium with many spiral defects.
We found no clear relation between the dimension correlation
length~$\xi_\delta$ and the widely used two-point
length~$\xi_2$; however, the numerical similarity of these
lengths supports a previous conjecture that 
$\xi_\delta \approx \xi_2$
in data taken on fibrillating hearts
\cite{Bayly93}. 
The mean Lyapunov dimension per defect of 3 to 7
suggests that defects are more complicated than
in the defect-turbulent regime of the
complex Ginzburg-Landau equation~\cite{Egolf97}, 
and that the dynamics of excitable media 
with even a few defects may be quite high-dimensional.  
This high-dimensionality in turn suggests that it will be
difficult to stabilize such states by small variations of
parameters~\cite{control-refs}, and may explain  
why some previous attempts to analyze the dynamics of
fibrillation with low-dimensional time series embedding
techniques have been inconclusive~\cite{embedding-refs}.
The unusual nature of the spiral-wave 
breakup in this medium leaves to future work the 
important question of whether the results obtained
here apply to other excitable media, including fibrillating
ventricles.

We thank A.~Karma, M.~B\"ar, P.~Bayly, and S.~Zoldi for
useful discussions, as well as two anonymous peer reviewers for
their useful comments on our manuscript. 
This work was supported by an NSF
Pre-Doctoral Research Fellowship, by NSF grants
NSF-DMS-93-07893 and NSF-CDA-92123483-04, and by DOE grant
DOE-DE-FG05-94ER25214.

\newpage

\begin{figure}   
\caption{Density plot at time $t=500$ of the slow field
$v(t,x,y)$ for a spatiotemporal chaotic state with
$31$~spiral defects present. Dark and light regions
correspond to values less or greater than the
value~$v^*=0.484$ corresponding to the unstable fixed point;
the field values span the range $v \in [0,a-b]$.  Parameter
values were $\epsilon = 0.074$, $a = 0.84$, $b = 0.07$, $L =
50$, $\Delta{x} =0.5$ and~$\Delta t = 0.0037$. }
\label{fig:pattern}
\end{figure}

\begin{figure}   
\caption{Fraction $f$ of $100$ random 
initial conditions still
exhibiting non-periodic dynamics after a given time.
({\bf a}) For biperiodic boundary conditions with cutoff time
$T_{np}=100$ (circles), 
$1000$ (squares), 
or any larger value, 
the transition has the same form, with systems of 
side length $L > 25$ nearly always exhibiting sustained
dynamics.
({\bf b}) For no-flux boundary conditions with cutoff time
$T_{np}=100$ (circles) and $T_{np}=1000$ (squares),
we see that the median transient time depends on the cutoff.
Comparison of these graphs shows that dynamics 
are substantially less likely 
to be sustained for a given time with no-flux boundary
conditions. The parameters used were the same as in
Figure 1.}
\label{fig:transients}
\end{figure}

\begin{figure}   
\caption{Lyapunov dimension~$D$ versus system area~$A=L^2$ of
Eq. (\ref{eq:bar-model}) for the parameter values
of Figure~\ref{fig:pattern}. Extensive (linear) scaling is found
for two different boundary conditions, no-flux (squares)
and periodic (circles). Data for~$L \leq 25$ did not
exist since all initial conditions decayed quickly to the
uniform state.  The dimension extrapolates to zero for a
positive system size, so the ratio $D/L^2$ of the dimension
to system area asymptotes slowly to the dimension
density~$\delta$.  }
\label{fig:extensivity}
\end{figure}

\begin{figure}   
\caption{ ({\bf a}) Dimension correlation length
$\xi_\delta$ (circles) and two-point correlation length
$\xi_2$ (squares) for different $\epsilon$~values.  ({\bf
b}) Degrees of freedom per mean defect, $D/\langle N\rangle$
as a function of $\epsilon$.  This ratio increases steadily
with~$\epsilon$ above the transition to chaos at $\epsilon =
\epsilon_c$ (marked by the arrow), 
and varies little in the region $\epsilon >
0.095$.}
\label{fig:lengths}
\end{figure}

\centerline{\epsfysize=3in \epsfbox{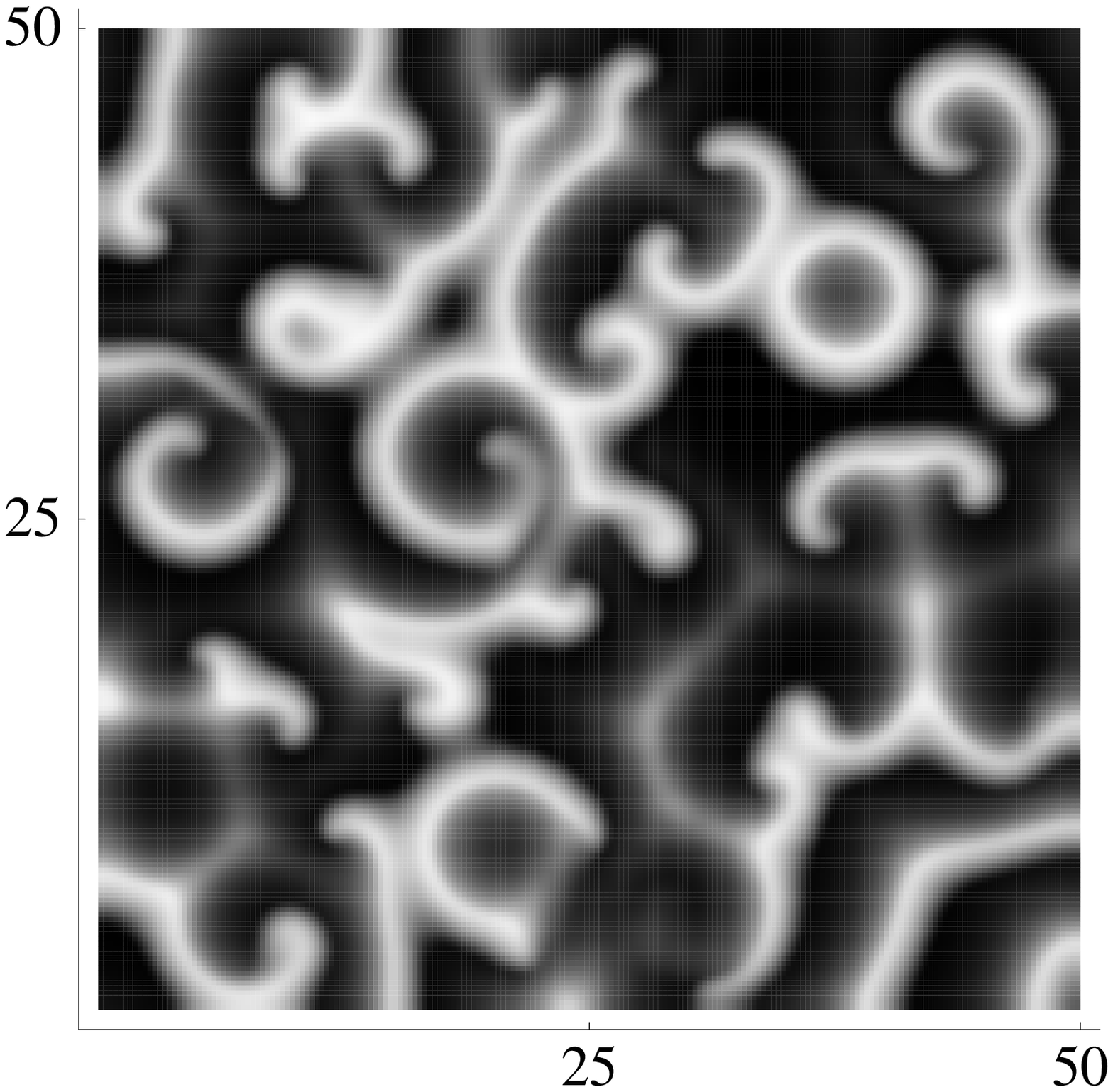}}
\centerline{\epsfysize=4in \epsfbox{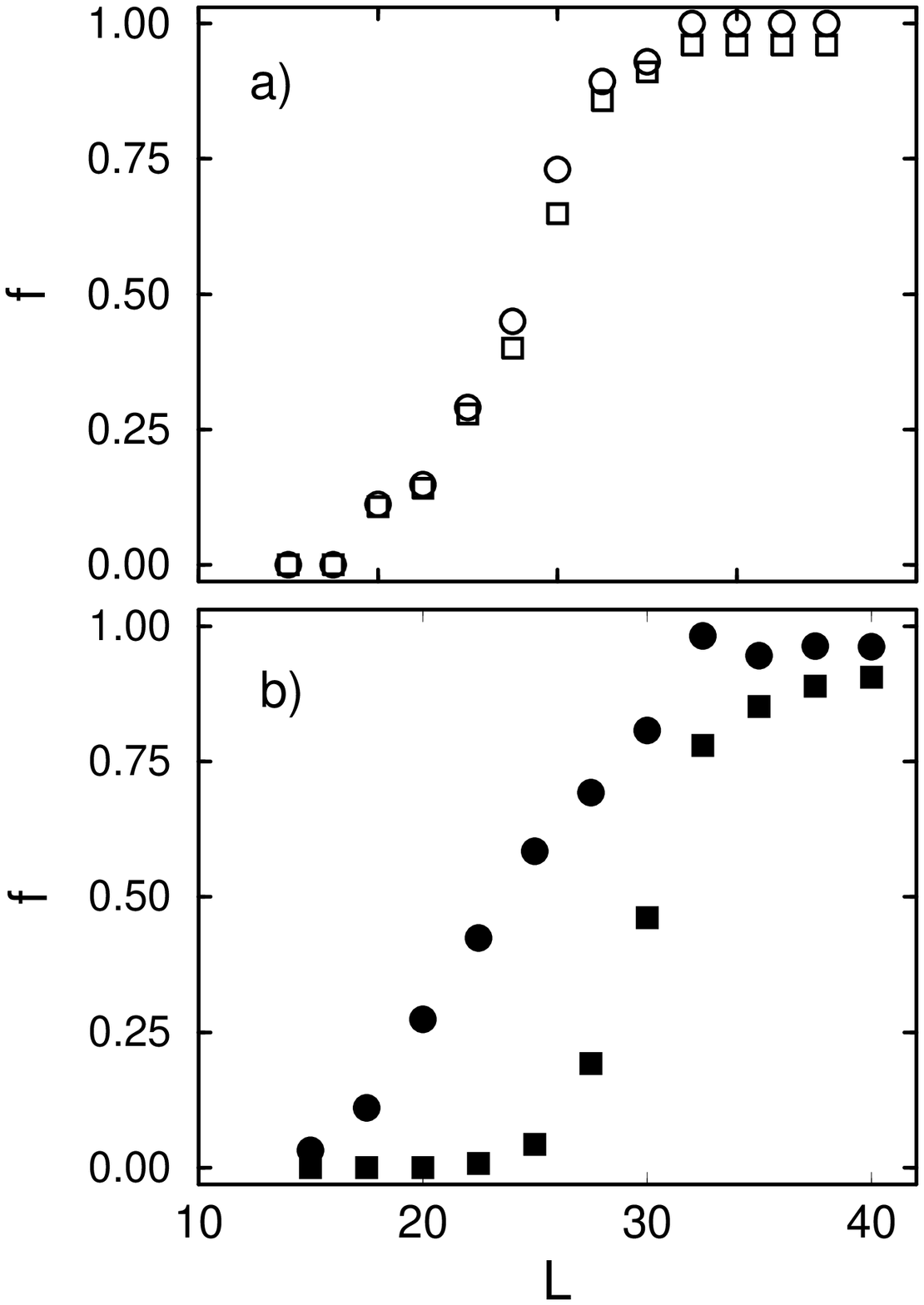}}
\centerline{\epsfysize=3in \epsfbox{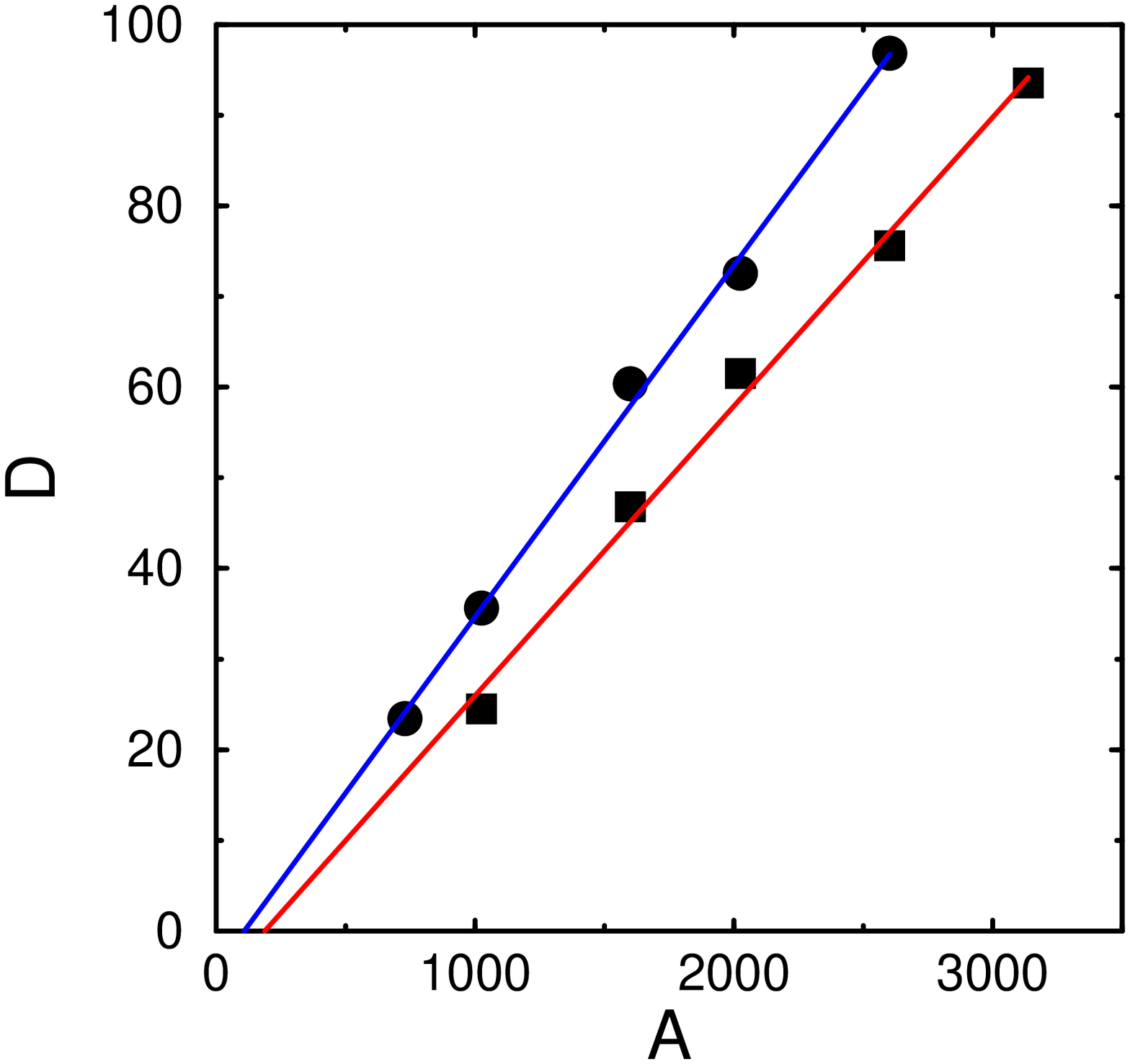}}
\centerline{\epsfysize=5in \epsfbox{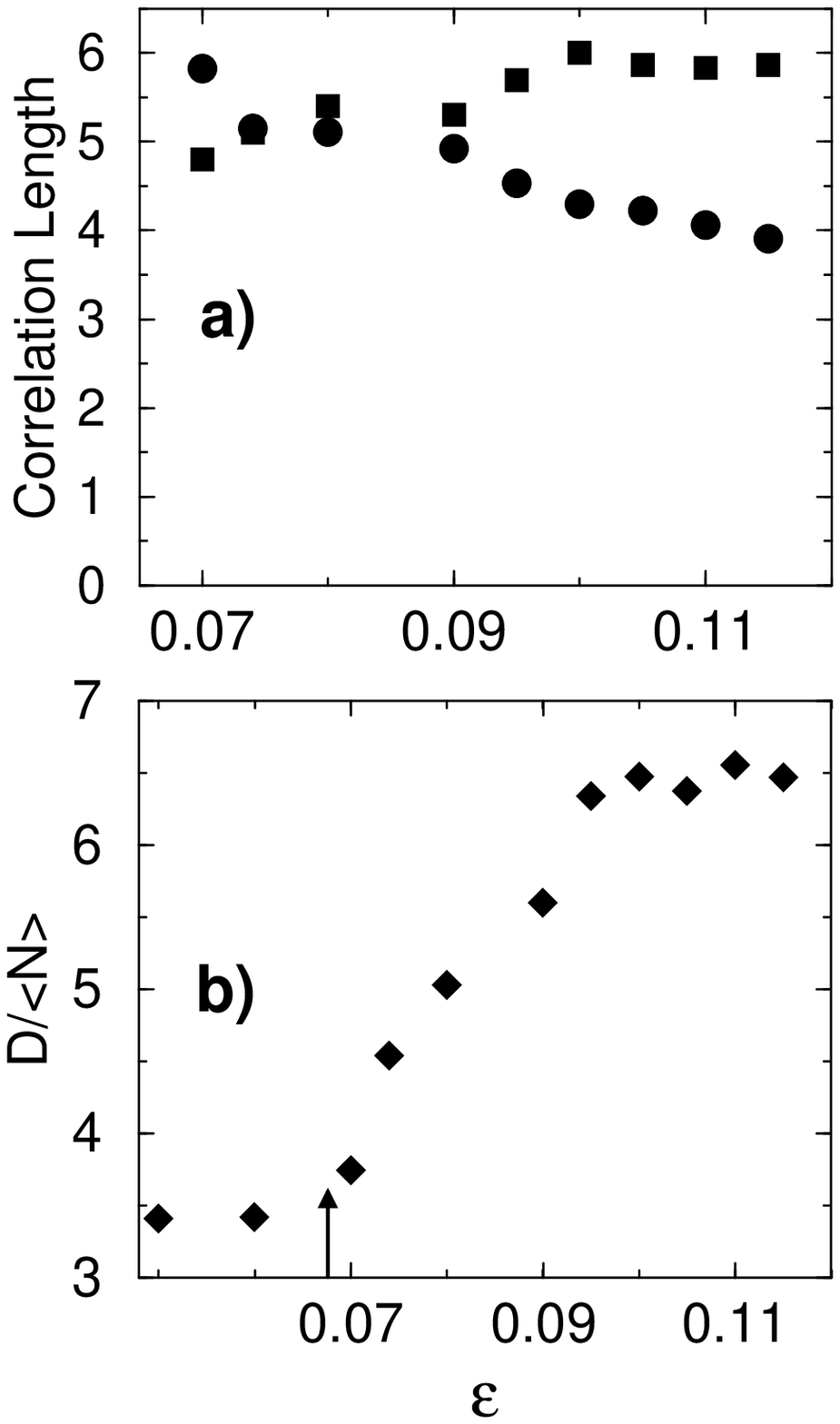}}

\end{document}